\documentclass[onecolumn,amsmath,amssymb,superscriptaddress,nofootinbib,11pt,a4paper]{revtex4}

\usepackage{titlesec}
\titleformat{\section}
  {\normalfont\Large\bfseries}  
  {\thesection}                 
  {1em}                         
  {}                            
\titleformat{\subsection}
  {\normalfont\large\bfseries}
  {\thesubsection}
  {1em}
  {}
\usepackage{xcolor}
\definecolor{mygreen}{RGB}{0, 155, 0}
\pdfoutput=1
\usepackage[T1]{fontenc}
\usepackage{CJK}
\usepackage{xcolor}
\usepackage{amsfonts}
\usepackage{epstopdf}
\usepackage{wrapfig} 
\usepackage{subcaption}
\usepackage{graphicx}  
\usepackage{dcolumn}   
\usepackage{bm}
\usepackage{float}
\usepackage[T1]{fontenc}
\usepackage{CJK}
\usepackage{xcolor}
\usepackage{amsfonts}
\usepackage{epstopdf}
\usepackage{wrapfig} 
\usepackage{subcaption} 
\usepackage{graphicx}  
\usepackage{dcolumn}   
\usepackage{bm}
\usepackage{float}
\usepackage[utf8]{inputenc}
\usepackage[T1]{fontenc}
\usepackage{booktabs}

\begin{document}

\title{Repetitive Penrose Process in Konoplya-Zhidenko Rotating Non-Kerr Black Holes}
\author{Xiao-Xiong Zeng}
\affiliation{College of Physics and Electronic Engineering, Chongqing Normal University, \\Chongqing 401331, China}

\author{Dong-Ping Su}
\affiliation{School of Intellignet Engineering, Chongqing City Management College, \\Chongqing 401331, China}

\author{Ke Wang\footnote{Electronic address: kkwwang2025@163.com  (Corresponding author)}}
\affiliation{School of Material Science and Engineering, Chongqing Jiaotong University, \\Chongqing 400074, China}

\begin{abstract}
{ This paper investigates the repetitive Penrose process in Konoplya-Zhidenko rotating non-Kerr black hole, exploring the influence of the deformation parameter on the repetitive Penrose process. After a brief review of the Konoplya-Zhidenko rotating non-Kerr black hole, we study the fundamental equations of the Penrose process in this spacetime, examine the iterative stopping conditions required for the repetitive Penrose process, and obtain the corresponding numerical results. It is concluded  that, in addition to previously observed phenomena, under the same decay radius, a larger initial dimensionless deformation parameter $\hat{\eta}$ leads to greater values of the energy return on investment and energy utilization efficiency, particularly at higher decay radii. Furthermore, a smaller initial $\hat{\eta}$ results in a larger maximum value of the energy return on investment. For energy utilization efficiency, the initial $\hat{\eta}$ should take an intermediate value to maximize its peak. Additionally, we find that a larger initial $\hat{\eta}$ corresponds to a smaller maximum value of the extracted energy.
}
\end{abstract}

\maketitle
\vspace{1ex}

 \newpage
\section{Introduction}
Black holes, as celestial bodies predicted by Einstein's theory of gravity, play a crucial role in numerous high-energy astrophysical phenomena, such as active galactic nuclei \cite{13,14,15,16}, gamma-ray bursts \cite{17,18,19}, and ultraluminous X-ray binaries \cite{20}, all of which release vast amounts of energy. According to general relativity, rotating black holes store enormous extractable energy within them. Extracting energy from black holes has always been a topic of interest for physicists, and exploring the mechanisms of energy extraction from black holes helps reveal the fundamental principles behind the generation of high-energy astrophysical phenomena.

The most famous mechanism for extracting energy from a black hole is the Penrose process \cite{10}, which extracts energy from a rotating black hole through particle splitting. An incident particle enters the ergosphere and splits into two parts. The part with negative energy falls into the black hole across the event horizon, while the other part escapes to infinity, carrying more energy than the original particle. However, this process requires that the relative three-dimensional velocity between the split particles must exceed half the speed of light \cite{11,12}, which is relatively difficult to achieve in reality. Penrose's pioneering work has inspired physicists to investigate other energy extraction mechanisms \cite{21,22,23,24,25,26,27,28,29,38}. Recently, Ruffini et al. \cite{30} imposed turning point conditions on the trajectories of particles in the Kerr black hole within the equations of motion of the original Penrose process, also achieving energy extraction. Furthermore, they proposed a repetitive Penrose process \cite{6} (for related early studies, see \cite{31}). The study found that in the repetitive Penrose process, not all of the extractable energy of the Kerr black hole can be extracted; the change in the extractable energy is primarily converted into irreducible mass. This occurs because after each Penrose process, the mass and spin of the new black hole must be used for energy extraction, while also considering the new irreducible mass. The repetitive Penrose process is nonlinear, and the irreducible mass also increases nonlinearly. The repetitive Penrose process proposed by Ruffini has now been extended to the repetitive electro-Penrose process \cite{5}, the Kerr-de Sitter black hole \cite{4}, and the accelerating Kerr black hole \cite{7}. In the repetitive electro-Penrose process, similarly, not all of the electrical energy of the Reissner-Nordström black hole can be depleted \cite{5}. In the Kerr-de Sitter and accelerating Kerr black holes, besides similar phenomena observed previously, the energy extraction capabilities of these black holes are, in some cases, stronger than that of the Kerr black hole.

The Konoplya-Zhidenko rotating non-Kerr metric deviates from the Kerr spacetime to a non-negligible extent. Studies have found that the Kerr metric with these deviations can also produce the same black hole ringing frequencies \cite{1}, providing an opportunity to explore theories beyond general relativity. Furthermore, investigations into quasi-periodic oscillation constraints \cite{32} and iron line spectroscopy \cite{33} further indicate that realistic astrophysical black holes can be described by the Konoplya-Zhidenko rotating non-Kerr metric. Therefore, it is necessary to further explore the characteristics of this rotating non-Kerr metric through various processes. Currently, the properties of the Konoplya-Zhidenko rotating non-Kerr black hole have been extensively studied \cite{34,2,8,35,36,37,39,40}. In this paper, we aim to further investigate the repetitive Penrose process in the Konoplya-Zhidenko rotating non-Kerr black hole and study the influence of the deformation parameter on this process. The results indicate that, in addition to previously observed phenomena, under the same decay radius, a larger initial dimensionless deformation parameter $\hat{\eta}$ leads to greater values of the energy return on investment and energy utilization efficiency, particularly at higher decay radii. Moreover, a smaller initial $\hat{\eta}$ results in a larger maximum value of the energy return on investment. For energy utilization efficiency, the initial $\hat{\eta}$ should take an intermediate value to maximize its peak. A larger initial $\hat{\eta}$ shifts the extracted energy curve to the right and results in a smaller maximum value of the extracted energy.

The remainder of this paper is organized as follows. In Section 2, we will introduce the Penrose process in the Konoplya-Zhidenko rotating non-Kerr black hole. In Section 3, we will discuss the iterative stopping conditions for this process. In Section 4, we will investigate the repetitive Penrose process in the Konoplya-Zhidenko rotating non-Kerr black hole, and in particular, we will explore the influence of the deformation parameter on this process. We present our conclusions in Section 5. Throughout this paper, we will use natural units ($c=G=1$).

\section{The Penrose Process in the Konoplya-Zhidenko Rotating Non-Kerr Black Hole}
In the Boyer-Lindquist coordinates, the Konoplya-Zhidenko rotating non-Kerr metric is given by \cite{1}
\begin{equation}
ds^{2} = -\left(1 - \frac{2Mr + \frac{\eta}{r}}{\rho^{2}}\right)dt^{2} + \frac{\rho^{2}}{\Delta} dr^{2} + \rho^{2}d\theta^{2} + \frac{A}{\rho^{2}}\sin^{2}\theta d\phi^{2} - \frac{2\left(2Mr + \frac{\eta}{r}\right)a\sin^{2}\theta}{\rho^{2}} dtd\phi,
\end{equation}
where the relevant metric functions are
\begin{equation}
\rho^{2} = r^{2} + a^{2}\cos^{2}\theta , \quad\Delta = r^{2} - 2Mr + a^{2} - \frac{\eta}{r}, \quad A = (r^{2} + a^{2})^{2} - a^{2}\Delta \sin^{2}\theta .
\end{equation}
Here, $M$ is the mass of the black hole, and $a$ is its spin. The parameter $\eta$ is the deformation parameter; when $\eta=0$, the metric reduces to the Kerr metric. The event horizon of the black hole is determined by $\Delta=0$, and the boundary of the ergosphere is determined by $g_{tt}=0$. For a detailed discussion of the event horizon and the ergosphere, see Ref. \cite{2}. In this paper, to compare with the Kerr black hole, we set the initial spin of the black hole to $a=M$ and consider particle motion in the equatorial plane. In this case, the ergosphere of the black hole lies between $r_+$ and $r_E$, with $\eta \geq 0$ \cite{2}. Here, the event horizon is
\begin{equation}
{r_+} = \frac{1}{3}\Bigg[2M + \frac{2^{1 / 3}(4M^2 - 3a^2)}{B^{1 / 3}} + \frac{B^{1 / 3}}{2^{1 / 3}}\Bigg],\label{3}
\end{equation}
where
\begin{equation}
B = 16M^3 +27\eta -18a^2 M + \sqrt{(16M^3 + 27\eta - 18a^2 M)^2 - 4(4M^2 - 3a^2)^3}.
\end{equation}
And the boundary of the ergosphere is
\begin{equation}
{r_E} ={\frac{1}{3}\bigg[2M + \frac{2^{1 / 3}(4M^{2} - 3a^{2}\cos^{2}\theta)}{C^{1 / 3}} +\frac{C^{1 / 3}}{2^{1 / 3}}\bigg],}
\end{equation}
where
\begin{equation}
C = 16M^{3} + 27\eta -18a^{2}M\cos^{2}\theta +\sqrt{(16M^{3} + 27\eta -18a^{2}M\cos^{2}{\theta})^{2} - 4(4M^{2} - 3a^{2}\cos^{2}{\theta})^{3}}.
\end{equation}

The surface area of the event horizon of the black hole is
\begin{equation}
S = \int_0^\pi \int_0^{2\pi} \sqrt{g_{\theta\theta} g_{\phi\phi}} \, d\phi d\theta =4\pi (r_+^2 + a^2).
\end{equation}
The irreducible mass of the black hole is \cite{3}\footnote{For another case regarding the irreducible mass, see the Supplementary Material.}
\begin{equation}
M_{irr} = \sqrt{\frac{S}{16\pi}}=\sqrt{\frac{r_+^2 + a^2}{4}}.\label{8}
\end{equation}
Then the extractable energy can be expressed as
\begin{equation}
E_{extractable}= M-M_{irr}=M-\sqrt{\frac{r_+^2 + a^2}{4}}.\label{89}\textbf{}
\end{equation}
This is the maximum energy theoretically extractable from the Konoplya-Zhidenko rotating non-Kerr black hole. 

The Konoplya-Zhidenko rotating non-Kerr black hole is not a strict solution of gravitational theory but rather a spacetime obtained by analogy. When Konoplya and Zhidenko designed this metric, they have emphasized that the introduction of the deformation parameter only changes the horizon position, without altering the asymptotic behavior or topological structure of the spacetime, therefore, many quantities can be derived by analogy, including the Bekenstein-Hawking area law. When the deformation parameter is zero, the law reduces to the Kerr case.

What needs to be emphasized is that $\eta$ is merely a deformation parameter of the metric, it is not a conserved charge, not a field, not a gravitational correction term, or any such physical parameter. In the 4D  Einstein-Gauss-Bonnet (EGB) theory, $\alpha$ is a higher order curvature coupling in the Lagrangian, which modifies gravity itself, and consequently the entropy necessarily receives corrections. However $\eta$ is completely different from such physical quantities; it cannot be altered in any physical process, nor does it change the state of any physical process. It is merely an initial condition of the gravitational system. Therefore, the thermodynamics of the KZ black hole should be similar to that of the Kerr black hole. The deformation parameter only changes the horizon, does not modify the entropy, and does not alter the Bekenstein-Hawking area law. Hence, Equation \eqref{8} and Equation \eqref{89}  are reasonable.

\begin{figure}[!h]
  \centering
  \includegraphics[width=0.9\linewidth]{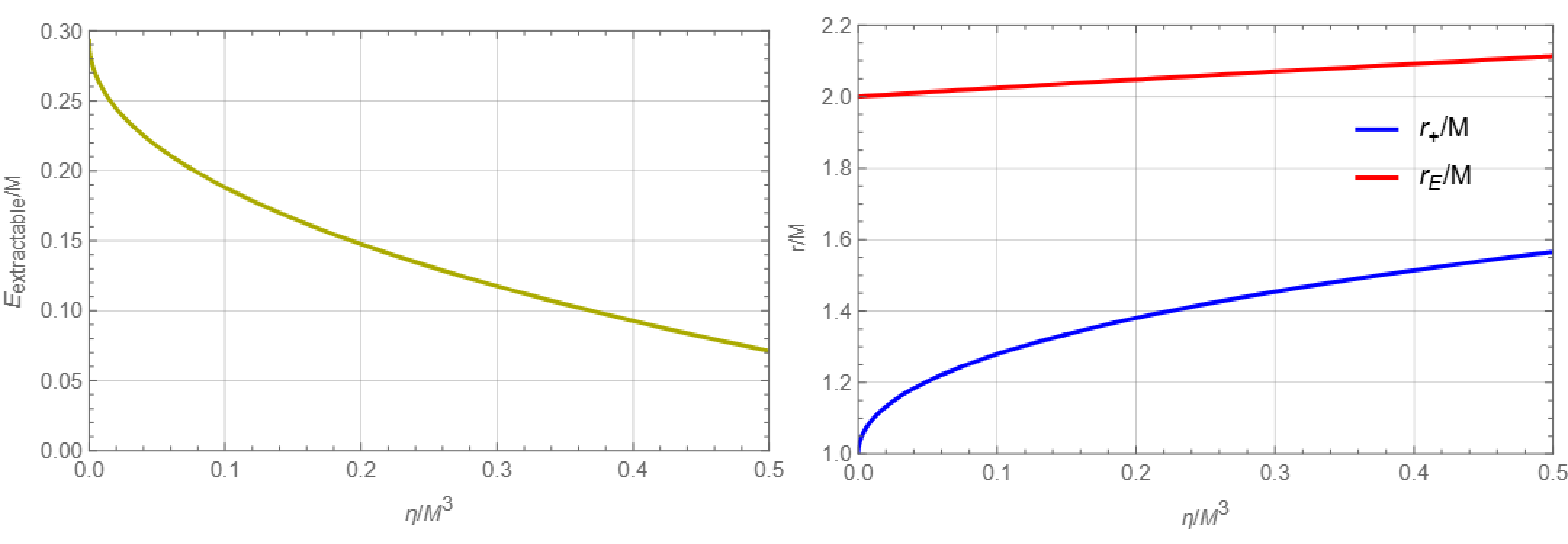}
  \caption{Variation of $E_{extractable}/M$, $r_{+}/M$, and $r_{E}/M$ with $\hat\eta$ for $a=M$.}
  \label{fig:1}
\end{figure}

We plot the maximum extractable energy, as well as the event horizon and ergosphere boundary for $a=M$, as functions of $\hat\eta=\eta/M^3$ in Fig. \ref{fig:1}.
It can be seen that as $\hat\eta$ increases, the event horizon and ergosphere boundary of the black hole for $a=M$ gradually increase, while the maximum extractable energy gradually decreases.

The fundamental equations of the Penrose process are given by the conservation of four-momentum. The conservation equations for energy, angular momentum, and radial momentum can be expressed as
\begin{equation}
\hat{E}_{0}=\tilde{\mu}_1\hat{E}_1+\tilde{\mu}_2\hat{E}_2, 
\end{equation}
\begin{equation}
\hat{p}_{\phi {0}}=\tilde{\mu}_1\hat{p}_{\phi 1}+\tilde{\mu}_2\hat{p}_{\phi 2},
\end{equation}
\begin{equation}
\hat{p}_{r0}=\tilde{\mu}_{1}\hat{p}_{r1}+\tilde{\mu}_2\hat{p}_{r2},
\end{equation}
where
\begin{equation}
\tilde{\mu}_{i}=\mu_i/\mu_0,\quad\hat{E}_i={E}_i/{\mu}_i,\quad\hat{p}_{\phi i}=p_{\phi i}/(\mu_iM),\quad\hat{p}_{ri}=p_{ri}/\mu_i,\quad i\in0,1,2. 
\end{equation}
Here, $\mu_{i} $ is the mass of particle $i$. The effective potential for a particle undergoing radial motion in the equatorial plane is \cite{4}
\begin{equation}
\hat{V}^\pm_{i} = \frac{ g^{\phi t} \hat{p}_{\phi i} M \mp \sqrt{ (g^{\phi t})^2 \hat{p}_{\phi i}^2 M^2 - g^{tt} ( g^{\phi\phi} \hat{p}_{\phi i}^2 M^2 + 1 ) } }{ g^{tt} }.
\end{equation}
We are interested in the optimal condition for maximum energy extraction, where the radial momenta of the three particles need to be zero at the decay location, with all three particles at their respective turning points, i.e., $\hat{E}_i = \hat{V}_i^+$. For specific reasons, see Ref. \cite{4}. Under these conditions, assuming $\hat{E}_0$, $\hat{p}_{\phi 1}$, and $\nu = \mu_2/\mu_1$ are known quantities, the fundamental equations of the Penrose process have an analytic solution \cite{4}
\begin{equation}
\hat{p}_{\phi 0} =\frac{ g^{\phi t} \hat{E}_0 +\sqrt{ (g^{\phi t})^2 \hat{E}_0^2 - g^{\phi\phi} (1 + g^{tt} \hat{E}_0^2) } }{ M g^{\phi\phi} },
\end{equation}
\begin{equation}
\hat{E}_1 = \frac{ g^{\phi t} \hat{p}_{\phi 1} M - \sqrt{ (g^{\phi t})^2 \hat{p}_{\phi 1}^2 M^2 - g^{tt} ( g^{\phi\phi} \hat{p}_{\phi 1}^2 M^2 + 1 ) } }{ g^{tt} },\label{16}
\end{equation}
\begin{equation}
\tilde{\mu}_1= \frac{\hat{E}_0\hat{E}_1g^{tt} - \hat{E}_1g^{\phi t}M\hat{p}_{\phi 0} - \hat{E}_0g^{\phi t}M\hat{p}_{\phi 1} + g^{\phi\phi}M^2\hat{p}_{\phi 0}\hat{p}_{\phi 1} + \sqrt{D}}{\hat{E}_1^2g^{tt} - 2\hat{E}_1g^{ \phi t}M\hat{p}_{\phi 1} + g^{\phi\phi}M^2\hat{p}_{\phi 1}^2 + \nu^2},
\end{equation}
\begin{equation}
\hat{p}_{\phi 2}=\frac{\hat{p}_{\phi 0}}{\tilde{\mu}_2}-\frac{\hat{p}_{\phi 1}}{\nu},\quad\hat{E}_2 =\frac{\hat{E}_0}{\tilde{\mu}_2}-\frac{\hat{E}_1}{\nu}, 
\end{equation}
where
\begin{equation}
\begin{aligned}
D= & - g^{tt} g^{\phi\phi} M^2 \hat{E}_1^2 \hat{p}_{\phi 0}^2  + (g^{t\phi})^2 M^2 \hat{E}_1^2 \hat{p}_{\phi 0}^2  - g^{tt} g^{\phi\phi} M^2 \hat{E}_0^2 \hat{p}_{\phi 1}^2  + (g^{t\phi})^2 M^2 \hat{E}_0^2 \hat{p}_{\phi 1}^2 \\
& - 2 (g^{t\phi})^2 M^2 \hat{E}_0 \hat{E}_1 \hat{p}_{\phi 0} \hat{p}_{\phi 1} + 2 g^{tt} g^{\phi\phi} M^2 \hat{E}_0 \hat{E}_1 \hat{p}_{\phi 0} \hat{p}_{\phi 1} + 2 g^{t\phi} M \hat{E}_0 \hat{p}_{\phi 0} \nu^2  - g^{tt} \hat{E}_0^2 \nu^2   \\
& - g^{\phi\phi} M^2 \hat{p}_{\phi 0}^2 \nu^2.
\end{aligned}
\end{equation}
After each energy extraction, the remaining mass and angular momentum of the black hole are
\begin{equation}
M_n=M_{n-1}+\Delta M_{n-1}=M_{n-1}+\hat{E}_{1,n-1}\mu_{1,n-1},\quad L_n=L_{n-1}+\hat{p}_{\phi 1}\mu_{1,n-1}M_{n-1},
\end{equation}
where
\begin{equation}
L_0=\hat{a}_0M_0^2.
\end{equation}
This leads to corresponding changes in $\hat{a}=a/M$ and $\hat{\eta}$, namely
\begin{equation}
\Delta\hat{a}_{n-1}=\frac{L_n}{M_n^2}-\frac{L_{n-1}}{M_{n-1}^2},\quad\Delta\hat{\eta}_{n-1}=\eta/M^3_n-\eta/M^3_{n-1}.\label{22}
\end{equation}
It is important to emphasize that during the repetitive Penrose process, we adopt a strong assumption that the deformation parameter $\eta$ remains constant (i.e., the intrinsic deviation of the black hole from the Kerr metric does not change as energy is extracted). The dimensionless deformation parameter 
$
\hat{\eta}_n = \frac{\eta}{M_n^3}
$
is updated iteratively. This assumption underlies the iteration rules presented in \eqref{22}.
Simultaneously, according to equations \eqref{3} and \eqref{8}, the event horizon and irreducible mass will also change. The change in the extractable energy is
\begin{equation}
\Delta E_{ extractable,{n-1}}=\Delta M_{n-1}-\Delta M_{ irr,{n-1}}.\,\,\,\,
\end{equation}
During this process, the extracted energy is
\begin{equation}
E_{ extracted,n}=M_{0}-M_{n}.\,\,\,
\end{equation}
The energy return on investment, defined as the ratio of the extracted energy to the total energy of all incident particles from infinity, is given by \cite{5}
\begin{equation}
\xi_{n}=E_{ extracted,n}/(nE_{0}).\,\,\,
\end{equation}
The energy utilization efficiency, defined as the ratio of the extracted energy to the difference between the initial and final extractable energy, is given by \cite{5}
\begin{equation}
\Xi_n=E_{ extracted,n}/(E_{ extractable,0}-E_{ extractable,n}).\,\,\,
\end{equation}
The above formulas serve as important parameters for measuring the effectiveness of energy extraction.

\section{Iterative Stopping Conditions}
\begin{figure}[!h]
  \centering
  \includegraphics[width=\linewidth]{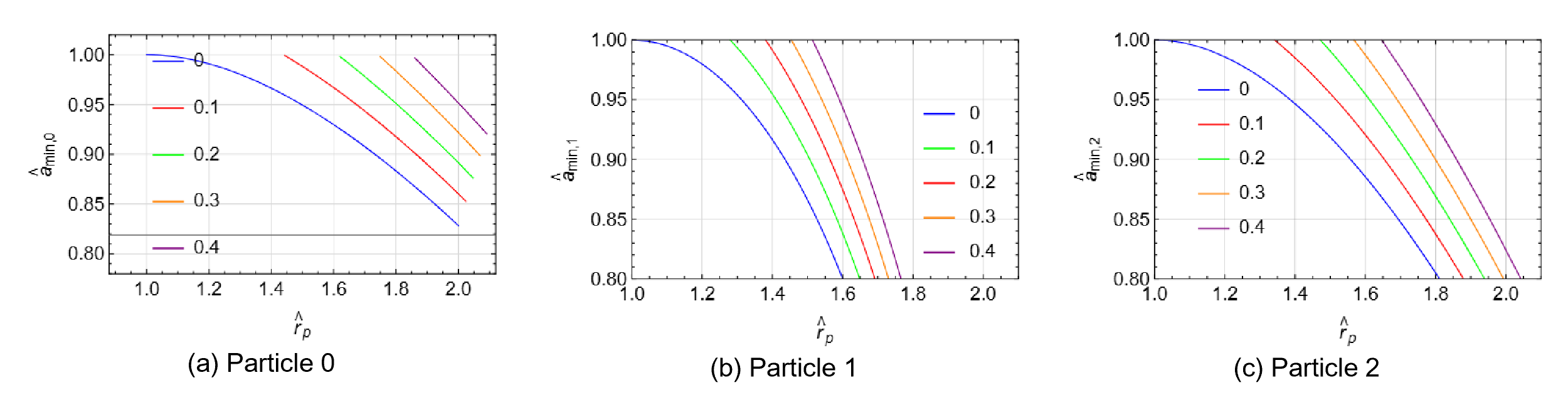}
  \caption{Variation of the minimum spin lower limits with the decay radius $\hat r_p$ for (a) particle 0, (b) particle 1, and (c) particle 2 under different $\hat{\eta}$.}
\label{fig:2}
\end{figure}

In the repetitive energy extraction process, the aforementioned iteration cannot proceed indefinitely; it must satisfy certain conditions, as described in \cite{4,6,7}. For example, first, the mass deficit must satisfy
\begin{equation}
1-\tilde{\mu}_1-\tilde{\mu}_2>0.\label{27}
\end{equation}
Second, during the iteration process, it must hold that $\hat{E}_{1}<0$. Third, for each iteration, it must hold that $E_{{extractable},n}>0$. Fourth, for each iteration, it must hold that the irreducible mass cannot decrease. The final condition is that the turning points of particles 0 and 2 must lie to the right of the peak of the effective potential, while the turning point of particle 1 must lie to the left of the peak of the effective potential. The corresponding limiting case occurs when the classical turning point of each particle is exactly at the peak of their respective effective potential, i.e.,
\begin{equation}
\hat{V}_i^{+}(\hat{r}_p)=\hat{E}_i, \quad{d}\hat{V}_i^+/{ d}\hat{r}|_{\hat{r}=\hat{r}_p}=0,
\end{equation}
where $\hat{r}_p={r}_p/M$ is the dimensionless decay radius. If $\hat{E}_0=1$, the minimum spin lower limit for stopping the iteration at this point is controlled by particle 0, and this minimum spin lower limit for particle 0 is located on its co-rotating marginally bound orbit. The angular velocity of a particle moving on a co-rotating Keplerian orbit in the Konoplya-Zhidenko rotating non-Kerr black hole is \cite{8}
\begin{equation}
\Omega = \frac{\sqrt{2Mr^{2} + 3\eta}}{a\sqrt{2Mr^{2} + 3\eta}+\sqrt{2r^{5}}}.
\end{equation}
Then the specific energy for a particle moving on a co-rotating Keplerian orbit is \cite{9}
\begin{equation}
\hat{\mathcal{E}} = -\frac{g_{tt} + g_{t\phi} \Omega}{\sqrt{-g_{tt} - 2g_{t\phi} \Omega - g_{\phi\phi} \Omega^2}}.\label{30}
\end{equation}
For a marginally bound orbit, $\hat{\mathcal{E}}=1$, and the minimum spin lower limit for particle 0 can be obtained by solving equation \eqref{30}. If $\hat{E}_0>1$, the minimum spin lower limit for stopping the iteration at this point is controlled by particle 2, and this minimum spin lower limit for particle 2 is located at the co-rotating photon sphere radius. The co-rotating photon sphere radius in the Konoplya-Zhidenko rotating non-Kerr black hole satisfies \cite{8}
\begin{equation}
(2r^{3} - 6Mr^{2} - 5\eta)^{2} - 8a^{2}(2Mr^{3} + 3\eta r) = 0.\label{31}
\end{equation}
Then the minimum spin lower limit for particle 2 can be obtained by solving equation \eqref{31}. For the turning point of particle 1 to exist, the discriminant under the square root in equation \eqref{16} must be positive, which corresponds to $\hat r_p$ being greater than the radius of the black hole's event horizon, with the critical case being $\hat r_p=\hat r_+$. In Fig. \ref{fig:2}, we plot the variation of the minimum spin lower limits for the three particles with $\hat r_p$ for different $\hat{\eta}$. For different values of $\hat{\eta}$, the ergosphere differs, so the range of the decay radius varies. We have neglected the very small portions of $\hat{a}_{min}$, which does not affect the trend of variation.
It can be seen from Fig. \ref{fig:2} that as the decay radius increases, the minimum spin lower limits for all three particles decrease. Under the same decay radius, as $\hat{\eta}$ increases, the minimum spin lower limits for the three particles increase.

\begin{figure}[!h]
  \centering
  \includegraphics[width=\linewidth]{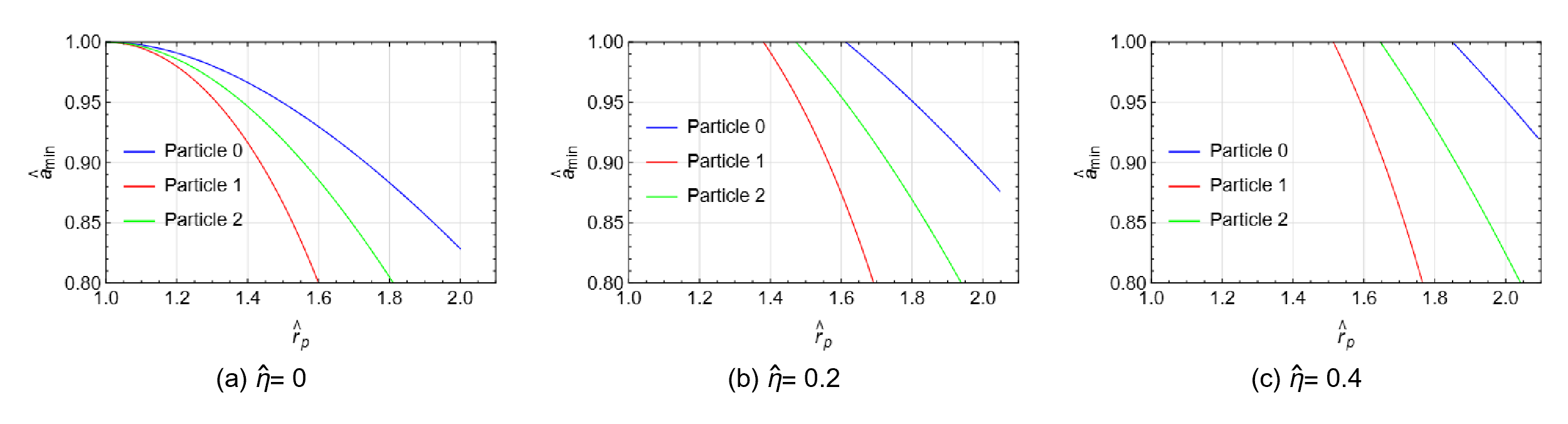}  
  \caption{Comparison of the minimum spin lower limits for the three particles for (a) $\hat{\eta}=0$, (b) $\hat{\eta}=0.2$, and (c) $\hat{\eta}=0.4$.}
  \label{fig:3}
\end{figure}

To determine the relative magnitudes of the minimum spin lower limits for the three particles, we plot Fig. \ref{fig:3}.
It can be seen that $\hat{a}_{{min},1} < \hat{a}_{{min},2} < \hat{a}_{{min},0}$. Therefore, the spin lower limit for stopping the iteration is controlled by particle 0. This is consistent with previous conclusions. It should be emphasized that the spin lower limit for stopping the iteration is not fixed, because with each iteration, the mass $M$ decreases and $\eta/M^3$ correspondingly increases. Thus, as can be seen from Fig. \ref{fig:2}(a), with each iteration, the spin lower limit for stopping the iteration increases slightly. In Fig. 2 of Ref. \cite{6}, Fig. 6 of Ref. \cite{4}, and Fig. 2 of Ref. \cite{7}, the variation of the effective potential of particle 0 with $\hat{r}$ is plotted, further explaining that the spin lower limit for stopping the iteration is controlled by particle 0.

\section{The Repetitive Penrose Process in the Konoplya-Zhidenko Rotating Non-Kerr Black Hole}
In this section, we investigate the repetitive Penrose process. The results of Ref. \cite{6} indicate that in the case of $\hat{E}_0>1$, the energy return on investment is lower than in the case of $\hat{E}_0=1$. Therefore, we choose $\hat{E}_0=1$ to maximize the energy return on investment. Following Ref. \cite{6}, we take $\hat{p}_{\phi 1}=-19.434$, $\nu = \mu_2/\mu_1=0.78345$, and we take $\mu_0 = 10^{-2}M_0$. We present the results in Table \ref{tab:1}, where the initial dimensionless deformation parameter is taken as $\hat{\eta}=0.2$. In this case, the ergosphere is located at $\hat{r}\in(1.3806,2.0477)$, and we take the decay radius as $\hat{r}_p=1.7$.
\begin{table}[htbp]
\centering
\small
\caption{The repetitive Penrose process for initial $\hat{\eta}=0.2$ and $\hat{r}_p=1.7$.}
\label{tab:1}
\begin{tabular}{ccccccccccc}
\hline
$n$ & $\frac{M_n}{M_0}$ & $\hat{a}_n$ & $\frac{\mu_{1,n}}{\mu_0}$ & $\frac{E_{1,n}}{\mu_0}$ & $\frac{E_{extractable,n}}{M_0}$ & $\frac{E_{extracted,n}}{M_0}$ & $\frac{M_{{irr,n}}}{M_0}$ & $\xi_n$ & $\Xi_n$ &$\hat{a}_{min,0,n}$ \\
\hline
0 & 1.000000 & 1.000000 & 0.022272 & -0.051092 & 0.147638 & 0.000000 & 0.852362 & 0.000000 & 0.000000 & 0.978071 \\
1 & 0.999489 & 0.996690 & 0.022155 & -0.051111 & 0.145360 & 0.000511 & 0.854129 & 0.051092 & 0.224285 & 0.978176 \\
2 & 0.998978 & 0.993398 & 0.022039 & -0.051131 & 0.143142 & 0.001022 & 0.855836 & 0.051102 & 0.227332 & 0.978281 \\
3 & 0.998467 & 0.990124 & 0.021923 & -0.051151 & 0.140982 & 0.001533 & 0.857485 & 0.051111 & 0.230358 & 0.978387 \\
4 & 0.997955 & 0.986868 & 0.021807 & -0.051171 & 0.138876 & 0.002045 & 0.859079 & 0.051121 & 0.233363 & 0.978492 \\
5 & 0.997443 & 0.983629 & 0.021691 & -0.051190 & 0.136821 & 0.002557 & 0.860622 & 0.051131 & 0.236350 & 0.978598 \\
6 & 0.996932 & 0.980409 & 0.021576 & -0.051210 & 0.134817 & 0.003068 & 0.862115 & 0.051141 & 0.239319 & 0.978704 \\
\hline
\end{tabular}
\end{table}
All data in Table \ref{tab:1} satisfy the iteration conditions. For example, according to the mass deficit formula \eqref{27}, we obtain $\tilde{\mu}_{1}<1/(1+\mu_{2}/\mu_{1})=0.56$. Furthermore, for each iteration, $\hat{E}_{1}<0$ and $E_{{extractable},n}>0$ are also satisfied, and the irreducible mass remains non-decreasing. The last column represents the minimum spin lower limit for stopping the iteration after each iteration, indicating that this lower limit increases slightly with each iteration. At $n=6$, the iteration has stopped; if we forcibly proceed to the 7th iteration, we have $\hat{a}_{7}=0.977207$ and $\hat{a}_{{min},0,7}=0.978811$, which no longer satisfy the iteration conditions.

It can be seen from Table \ref{tab:1} that a small portion of the decrease in extractable energy flows into the extracted energy, while the majority flows into the irreducible mass. According to the energy utilization efficiency, $23.9\%$ of the change in extractable energy is converted into extracted energy, and $76.1\%$ of the change in extractable energy is converted into irreducible mass. The results indicate that reducing the black hole's spin cannot extract all of the corresponding rotational energy, a limitation arising from the nonlinear increase in irreducible mass. Moreover, in the repetitive Penrose process, after the iteration terminates, the remaining extractable energy is $0.134817M$, suggesting that a considerable amount of energy remains to be extracted through other means. These results exhibit similarities to the Kerr black hole case \cite{6}.

\begin{figure}[!h]
  \centering
  \includegraphics[width=\linewidth]{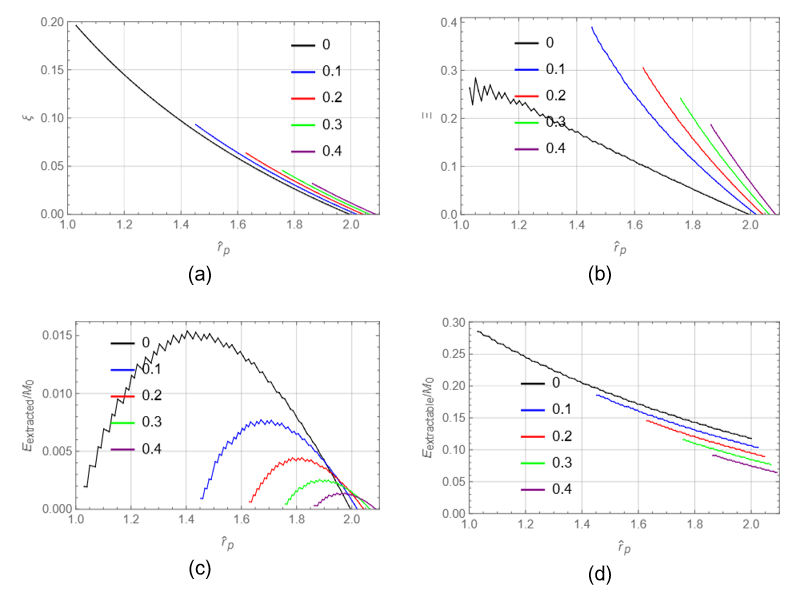}
  \caption{For different initial $\hat{\eta}$, after the termination of the repetitive Penrose process, variation with the decay radius $\hat r_p$ of (a): the energy return on investment $\xi$, (b): the energy utilization efficiency $\Xi$, (c): the extracted energy $E_{{extracted}}/{M_0}$, and (d): the extractable energy $E_{{extractable}}/{M_0}$. Each oscillation in the curves represents a different number of iterations, caused by the iteration conditions and reflecting the discrete nature of the iteration.}
  \label{fig:4}
\end{figure}

Finally, in Fig. \ref{fig:4}, we plot the variation with the decay radius $\hat r_p$ of the energy return on investment $\xi$, the energy utilization efficiency $\Xi$, the extracted energy $E_{{extracted}}/{M_0}$, and the extractable energy $E_{{extractable}}/{M_0}$ after the termination of the repetitive Penrose process for different initial $\hat{\eta}$.
From Fig. \ref{fig:4}(a,b), it can be seen that under the same decay radius, a larger initial dimensionless deformation parameter $\hat{\eta}$ leads to greater values of the energy return on investment and energy utilization efficiency, particularly at higher decay radii. Furthermore, a smaller initial $\hat{\eta}$ results in a larger maximum value of the energy return on investment. For the energy utilization efficiency, the initial $\hat{\eta}$ should take an intermediate value to maximize its peak. From Fig. \ref{fig:4}(c), it can be seen that the extracted energy exhibits a downward-opening parabola; a larger initial $\hat{\eta}$ shifts the curve to the right, and the peak value of the extracted energy is smaller. From Fig. \ref{fig:4}(d), it can be seen that a larger initial $\hat{\eta}$ results in a smaller extractable energy.

It can be seen from Fig. \ref{fig:4} that the curve is not continuous but oscillatory. To explain this phenomenon, we take $\hat{\eta}=0$ and list in Table \ref{tab:xi_rp} the values of the energy utilization efficiency $\Xi$ and the corresponding iteration numbers for the decay radius $\hat{r}_p$ from 1.03 to 1.20 in steps of 0.01. The purpose of taking this set of data is that, under this condition, as shown in Fig. \ref{fig:4}(b), the curve oscillates particularly strongly and is easier to distinguish.
\begin{table}[htbp]
\centering
\caption{Values of energy utilization efficiency $\Xi$ and corresponding iteration numbers for decay radius $\hat{r}_p$ from 1.03 to 1.20 in steps of 0.01. The last column indicates when $\Xi$ increases compared to the previous row.}
\label{tab:xi_rp}
\begin{tabular}{ccc c}
\hline
$\hat{r}_p$ & $\Xi_n$ & $n$ &  \\
\hline
1.03 & 0.265175 & 1 &  \\
1.04 & 0.229494 & 1 &  \\
1.05 & 0.285791 & 2 & Oscillation \\
1.06 & 0.259022 & 2 &  \\
1.07 & 0.237700 & 2 &  \\
1.08 & 0.268522 & 3 & Oscillation \\
1.09 & 0.250497 & 3 &  \\
1.10 & 0.270890 & 4 & Oscillation \\
1.11 & 0.255284 & 4 &  \\
1.12 & 0.241464 & 4 &  \\
1.13 & 0.256095 & 5 & Oscillation \\
1.14 & 0.243637 & 5 &  \\
1.15 & 0.254640 & 6 & Oscillation \\
1.16 & 0.243281 & 6 &  \\
1.17 & 0.232817 & 6 &  \\
1.18 & 0.241317 & 7 & Oscillation \\
1.19 & 0.231592 & 7 &  \\
1.20 & 0.238271 & 8 & Oscillation \\
\hline
\end{tabular}
\end{table}
It can be seen from Table \ref{tab:xi_rp} that as the decay radius increases, the number of iterations changes, leading to oscillations in the curve. Each oscillation of the curve corresponds to a difference in the iteration number for adjacent decay radii, indicating the discrete nature of the iteration.

\section{Conclusion}
In this paper, we have investigated energy extraction via the repetitive Penrose process in the Konoplya-Zhidenko rotating non-Kerr black hole under the optimal conditions for maximum energy extraction. To compare with the Kerr black hole, we set the initial spin of the black hole to $a=M$ and considered particle motion in the equatorial plane. First, we provided a brief review of the Konoplya-Zhidenko rotating non-Kerr black hole, including its horizon and ergosphere. Second, we introduced the fundamental equations of the Penrose process in this spacetime. Then, we discussed the iterative stopping conditions required for the Penrose process. In particular, we plotted the variation of the minimum spin lower limits for particles 0, 1, and 2 with the decay radius $\hat r_p$ for different $\hat{\eta}$, and compared the minimum spin lower limits of the three particles, concluding that the spin lower limit for stopping the iteration is controlled by particle 0. Finally, we presented the corresponding numerical results. Similar to previous findings, reducing the black hole's spin cannot extract all of the corresponding rotational energy, a limitation arising from the nonlinear increase in irreducible mass. The difference lies in that, under the same decay radius, a larger initial dimensionless deformation parameter $\hat{\eta}$ leads to greater values of the energy return on investment and energy utilization efficiency, particularly at higher decay radii. Furthermore, a smaller initial $\hat{\eta}$ results in a larger maximum value of the energy return on investment. For the energy utilization efficiency, the initial $\hat{\eta}$ should take an intermediate value to maximize its peak. A larger initial $\hat{\eta}$ shifts the extracted energy curve to the right and results in a smaller maximum value of the extracted energy.

\noindent {\bf Acknowledgments}

\noindent
This work is supported by the National Natural Science Foundation of China (Grants Nos.
12375043, 12575069 ), and Chongqing Normal University Fund Project (Grants No. 26XLB001).

\noindent {\bf Conflict of Interest}

\noindent
The authors declare that they have no conflict of interest.

\end{document}